\documentclass{aa}
\usepackage{xspace}
\usepackage{natbib}
\usepackage{graphicx}
\usepackage{amsmath}
\usepackage{amssymb}
\usepackage{url}
\usepackage{txfonts}
\usepackage{rotating}

\newcommand{\ca}{\mbox{$\sim$}}
\newcommand{\kevnxs}{\ensuremath{\text{ke\kern -0.09em V}}}
\newcommand{\keV}{\kevnxs\xspace}
\newcommand{\err}[2]{\ensuremath{^{+#1}_{-#2}}\xspace}

\newcommand{\Msun}{\ensuremath{\mbox{M}_\odot}\xspace}
\newcommand{\Lsun}{\ensuremath{\mbox{L}_\odot}\xspace}
\newcommand{\xte}{\textsl{RXTE}\xspace}
\newcommand{\pca}{\textsl{PCA}\xspace}
\newcommand{\hexte}{\textsl{HEXTE}\xspace}

\newcommand{\nh}{\ensuremath{N_{\text{H}}}\xspace}
\newcommand{\ev}{e\kern -0.11em V\xspace}
\newcommand{\redchi}{\ensuremath{\chi^2_{{\text{red}}\xspace}}}
\newcommand{\gx}{GX\,301$-$2\xspace}
\newcommand{\wray}{Wray\,977\xspace}
\newcommand{\vela}{Vela~X-1\xspace}
\newcommand{\ecut}{\ensuremath{E_{\text{Cut}}}\xspace}
\newcommand{\efold}{\ensuremath{E_{\text{F}}}\xspace}
\newcommand{\ecyc}{\ensuremath{E_{\text{C}}}\xspace}
\newcommand{\sbf}{\ensuremath{\sigma_{\text{bf}}}}

\begin{document}
\title{The variable cyclotron line in \gx} 

\author{Ingo Kreykenbohm\inst{1,2} \and J\"orn Wilms\inst{3} \and Wayne
  Coburn\inst{4} \and Markus Kuster\inst{5} \and Richard E.
  Rothschild\inst{6} \and William A.  Heindl\inst{6} \and Peter
  Kretschmar\inst{5,2} \and R\"udiger Staubert\inst{1} }
\offprints{I. Kreykenbohm,\\e-mail: kreyken@astro.uni-tuebingen.de}

\institute{Institut f\"ur Astronomie und Astrophysik -- Astronomie,
  Sand 1, 72076 T\"ubingen, Germany \and INTEGRAL Science Data Centre,
  6 ch.\ d'\'Ecogia, 1290 Versoix, Switzerland \and Department of
  Physics, University of Warwick, Coventry CV4 7AL, U.K. \and Space
  Sciences Laboratory, University of California, Berkeley, Berkeley,
  CA, 94702-7450, U.S.A.  \and Max-Planck-Institut f\"ur
  extraterrestrische Physik, Giessenbachstr.~1, 85740 Garching,
  Germany \and Center for Astrophysics and Space Sciences, University
  of California at San Diego, La Jolla, CA 92093-0424, U.S.A.  }

\date {Received: 10 December 2003 / Accepted: 11 August 2004}

\abstract{We present pulse phase resolved spectra of the hypergiant
  high mass X-ray binary \gx. We observed the source in 2001 October
  with \xte continuously for a total on-source time of almost
  200\,ksec. We model the continuum with both, a heavily absorbed
  partial covering model and a reflection model.  In either case we
  find that the well known cyclotron resonant scattering feature
  (CRSF) at \ca35\,\keV is -- although present at all pulse phases --
  strongly variable over the pulse: the line position varies by 25\%
  from 30\,\keV in the fall of the secondary pulse to 38\,\keV in the
  fall of the main pulse where it is deepest. The line variability
  implies that we are seeing regions of magnetic field
  strength varying between $3.4\times 10^{12}$\,G and $4.2\times
  10^{12}$\,G. }

\maketitle

\section{Introduction}
\object{GX 301$-$2} (4U\,1223$-$62) is a High Mass X-ray Binary system
consisting of a neutron star accreting from the strong stellar wind of
the early type B-emission line star \wray. Due to difficulties in
classifying \wray, the system parameters are rather uncertain.  In a
detailed analysis, \citet{kaper95a}  classified \wray as a
B1\,Iae+ hypergiant at a distance of 5.3\,kpc; we will adopt these
values throughout this paper. If this classification 
is correct, \wray is one of the most luminous and massive
stars in our galaxy, with a luminosity of $1.3\times 10^6\,\Lsun$ and
a mass in excess of 48\,\Msun (best fit \ca75\,\Msun). With one of the
highest known wind mass loss rates in the Galaxy ($\dot{M} \sim
10^{-5}\,\Msun\,\text{yr}^{-1}$), the very slow wind ($v_\infty
=400\,\text{km}\,\text{s}^{-1}$) can easily feed the neutron star with
enough material to explain the observed X-ray luminosity of
$10^{37}\,\text{erg}\,\text{s}^{-1}$. We note that \citet{koh97a}
doubt these extreme parameters for \wray as they are only barely
compatible with their analysis, and argue that the earlier distance of
1.8\,kpc \citep{parkes80a} and a spectral type of B2\,Iae is correct,
with a correspondingly reduced luminosity of the Be primary.  Even with
these more moderate parameters, however, the system is unique. Note that the
following analysis does not depend on the detailed choice of system
parameters.

\begin{figure}
\includegraphics[width=1.0\columnwidth]{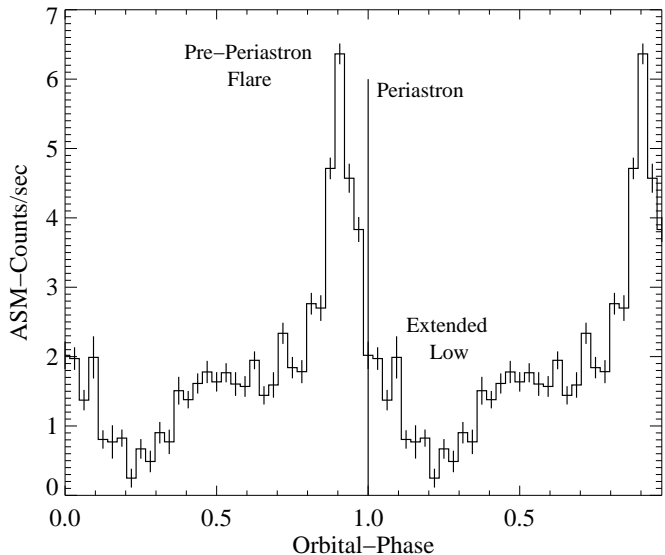}
\caption{Folded light curve using all available data (starting
  in 1996 until mid 2003) on \gx of the All Sky Monitor on board \xte.
  The light curve has been folded with the orbital period of 41.498\,d
  \citep{koh97a}. The periastron passage has been extrapolated based
  on the ephemeris of \citet{koh97a}. The flare shortly before the
  periastron passage is very evident. Note the extended low following
  the periastron passage which is probably due to the optical companion
  almost eclipsing the neutron star. For clarity the folded light
  curve is shown twice.}
\label{asmlight}
\end{figure}

\begin{figure}
\includegraphics[width=\columnwidth]{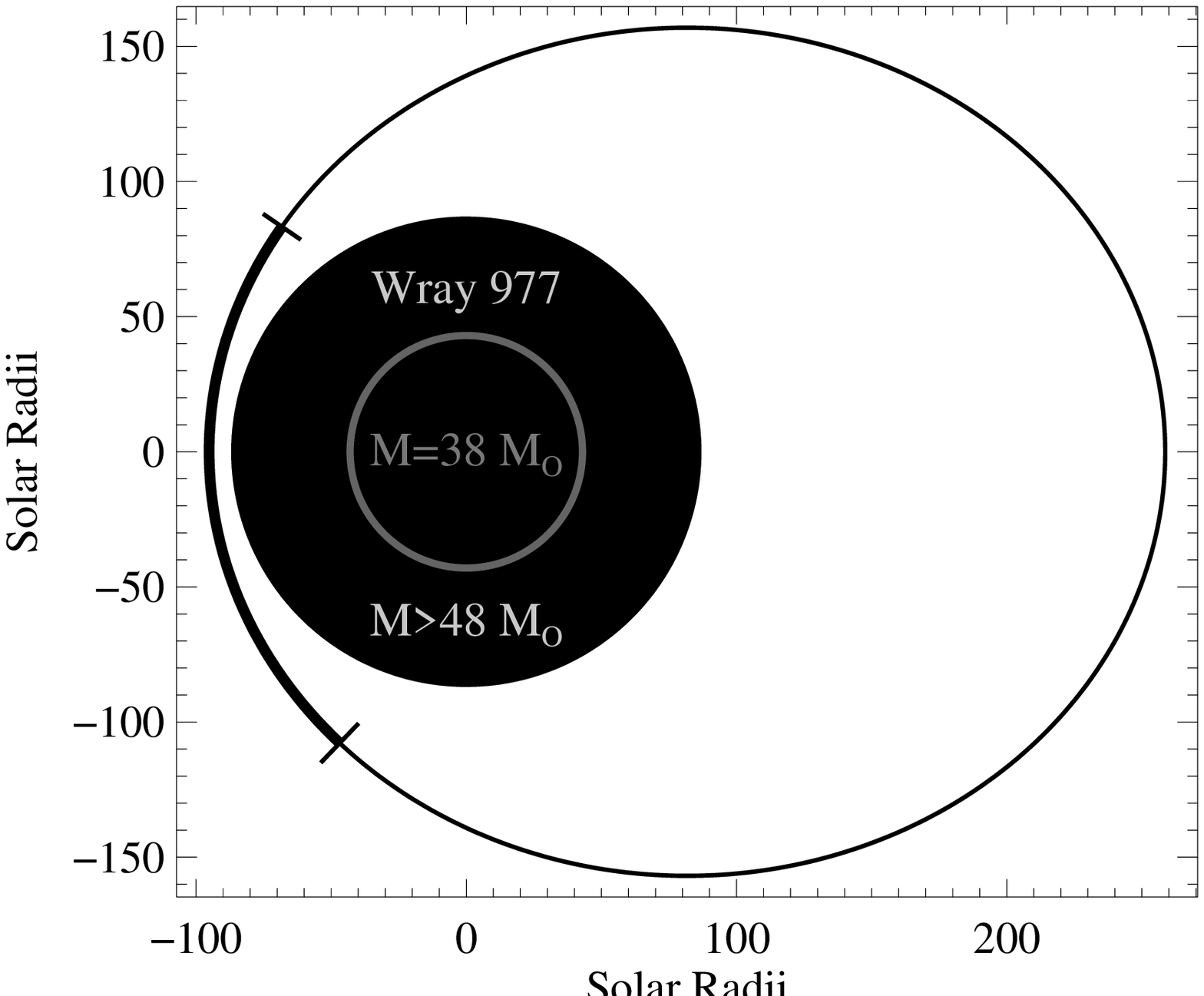}\vspace*{2mm}

\caption{Sketch of the system \gx/\wray based on the parameters of
  \citet{kaper95a}. The neutron star passes \wray at a distance of
  $\lesssim 0.1R_\star$ during the periastron passage. The time of the
  observation is marked by dashes and stronger line thickness,
  covering a significant part of the orbit due to the high velocity of
  the neutron star during periastron passage. The grey inner circle
  represents the size of \wray when using the old values of
  \citet{parkes80a} which are also used by \citet{koh97a}.}
\label{orbitsketch}
\end{figure}

The neutron star is in an eccentric ($e=0.462$) orbit with a period of
$41.498\pm0.002$\,days \citep{koh97a,sato86b}. Due to the violent wind
accretion the source is extremely variable -- X-ray luminosity changes
of a factor of five within one hour being common
\citep{rothschild87a}.  The underlying orbital variability, however,
always follows the same pattern (Fig.~\ref{asmlight}): Shortly before
periastron passage, the neutron star intercepts the gas stream from
\wray \citep{leahy91a,leahy02a}, resulting in an extended X-ray flare
during which the luminosity increases by a factor of \ca25 \citep[see
Fig.~\ref{asmlight} and][]{pravdo95a}.  During periastron passage, the
neutron star passes through the dense inner wind of \wray,
$\ca0.1\,R_\star$ above the stellar surface \citep[see][and
Fig.~\ref{orbitsketch}]{pravdo01a}, and the X-ray activity decreases
until it reaches a minimum after the periastron passage.  This
behavior is probably due to the neutron star being almost
eclipsed by \wray \citep{leahy02a} and an intrinsically lower
luminosity compared to the flare.  Following this minimum, the X-ray
luminosity increases slowly again over the orbit with a possible
second maximum near apastron, where the neutron star intercepts the
spiral shaped gas stream from \wray a second time \citep{leahy02a}.

\begin{figure}
\includegraphics[width=\columnwidth]{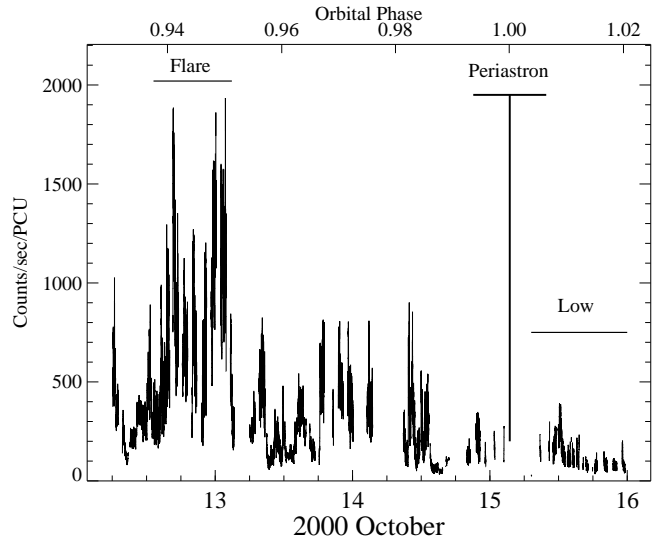}
\caption{X-ray light curve of the periastron  passage of \gx as
  observed with \xte. The count rate is given in counts/sec per
  Proportional Counter Unit (PCU). Note the pre-periastron flare, the
  low following the periastron passage, and the overall variability of
  the source.  The time of the periastron passage has been
  extrapolated using the ephemeris of \citet{koh97a}; the horizontal
  bar represents the uncertainty due to the uncertainties of the
  ephemeris. Our observation covered a significant part of the
  pre-periastron flare and the periastron passage itself.  During the
  flare, count rates are up to 20 times higher than in the low
  following the periastron passage.}
\label{lightcurve}
\end{figure}

Using the Rossi X-ray Timing Explorer (\xte) we observed \gx starting
in 2000 October (JD\,2451829.8 to JD\,2451868.3) covering the greater
part of the pre-periastron flare and the complete periastron passage
in one contiguous \ca200\,ksec long observation
(Fig.~\ref{lightcurve}). We used the same procedures as in our earlier
analysis of \xte data from \object{Vela~X-1} \citep{kreykenbohm02b}.
Due to the improved calibration of the Proportional Counter Array
(\pca), the systematic uncertainties associated with the \pca differ
with respect to our earlier work \citep[compare Table~\ref{systerror}
and][Table~1]{kreykenbohm02b}.  All analysis was done using
\textsl{FTOOLS 5.2} and \textsl{XSPEC 11.2.0au} \citep{arnaud96a}.
Phase resolved spectra were produced with a modified version of the
\textsl{FTOOL fasebin}, which properly accounts for the \hexte dead
time \citep{kreykenbohm02b}.

The remainder of this paper is structured as follows. In
Sect.~\ref{pulse} we discuss the pulse period and pulse profiles.
Sect.~\ref{spectra} discusses the different possible spectral models
and detailed fits to phase resolved spectra with special emphasis on
the behavior of the cyclotron line. We summarize our results in
Sect.~\ref{summary}.

\begin{table}
\caption{\label{systerror}Systematic errors applied to the \pca-data
  to account for uncertainties in the \pca calibration. We
  derived these values by fitting a two power law model simultaneously
  to a spectrum of a public \xte observation of the Crab. See
  \citet{wilms99b} and \citet{kreykenbohm02b} for a more detailed
  discussion of this procedure.}
\begin{tabular}{rrc}
\hline
\hline
Channels & Energy & Systematics \\
\hline
\phantom{4}0--\phantom{1}15& \phantom{0}2--\phantom{0}8\,\keV & 1.0\%\\
16--\phantom{1}39 & \phantom{0}8--18\,\keV & 0.5\%\\
40--\phantom{1}57 & 18--29\,\keV & 2.0\% \\
58--128 & 29--120\,\keV & 5.0\%\\
\hline
\end{tabular}
\end{table}

\section{Pulse profiles}
\label{pulse}
With a pulse period of \ca700\,sec \citep{swank76a}, \gx is one of the
slowest pulsars known. Since the first measurement of the period with
\textsl{Ariel-5} \citep{white76a}, it has undergone dramatic
variations: the period was stable at \ca700\,sec until 1984, then the
pulsar was spun up between 1985 and 1990 \citep{koh97a,bildsten97a} to
a period of \ca675\,sec. In 1993, the spin behavior reversed
\citep{pravdo01a} and the pulsar has since been spinning down almost
continously: we determined the pulse period of our observation (JD\,
2451830.8) to be 684.2\,sec (see below).  These long lasting spin-up
and spin-down trends are in favor of \gx being predominantly
a disk accreting system.  Similar to other HMXB systems like Vela~X-1,
however, the pulse to pulse variations are very strong and the overall
short term evolution of the pulse period is best described by a random
walk model as is typical for a wind accretor \citep{dekool93a}.

To obtain pulse profiles, the bin (\pca) or photon arrival (\hexte)
times were first barycentered and then corrected for the orbital
motion of \gx using the ephemeris of \citet{koh97a}. We used epoch
folding \citep{leahy83a} to determine the pulse period during the
periastron passage to be $684.2\pm0.2$\,sec, thus confirming that \gx
is continuing its spin down.

\begin{figure}
\includegraphics[width=\columnwidth]{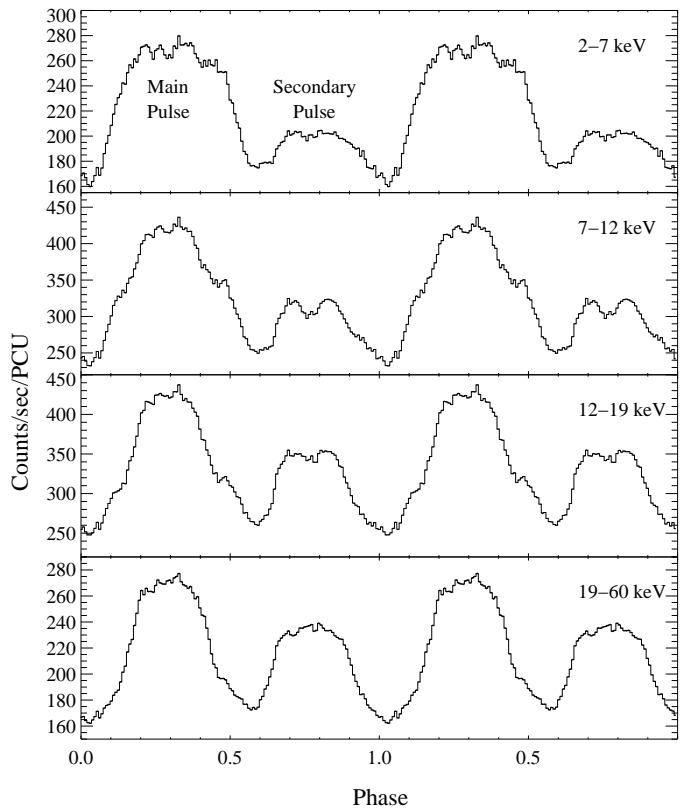}
\caption{Pulse profiles in four energy bands from data taken
  \emph{during} the pre-periastron flare (see Fig.~\ref{lightcurve}).
  For clarity the pulse profiles are shown twice. Note that error
  bars are shown, but are too small to be seen in print.}
\label{pulseprofiles}
\end{figure}

We used this period to create energy resolved pulse profiles. The long
period, the pulse to pulse variations, and the strong variability of
the source exacerbate the determination of a pulse profile. In our
case, \gx is brighter by a factor of more than ten during the
pre-periastron flare compared to the low during the actual periastron
passage (Fig.~\ref{lightcurve}).  We therefore created several sets of
energy resolved pulse profiles using \pca-data: For the first set we
used data taken during the flare only (see Fig.~\ref{pulseprofiles}),
while we used data from the low and directly after the pre-periastron
flare for the second and third set (Fig.~\ref{phase_desc}).  In
general, the average pulse profile of \gx is very well defined and
similar to that of \vela: it consists of two strong pulses at higher
energies evolving into a more complex shape with substructures below
12\,\keV. As one of the peaks is stronger in all energy bands than the
other, we designate this peak as the \emph{main pulse} and the weaker
peak as the \emph{secondary pulse}. The secondary peak is strongly
energy-dependent: it is very weak (compared to the main pulse) at
energies below 10\,\keV. At energies above \ca20\,\keV it is
significantly stronger, in fact, almost as strong as the main pulse.
At these higher energies, the profile is approximately sinusoidal;
however, the main pulse remains stronger throughout the energy band
covered here.  Also similar to \vela is that the long term pulse
profile is very stable. There are, however, significant short term
variations.

The interpretation of this energy dependence of the pulse profile is
not straightforward.  Above \ca20\,\keV, the sinusoidal pulse shape
can be attributed to contributions from the two magnetic poles of the
neutron star. It has been argued that anisotropic propagation of the
X-rays in the magnetized plasma of the accretion column
\citep{nagel81b} or a complicated configuration of the magnetic field
\citep{mytrophanov78a} could produce the complex pulse profile at
lower energies as well.  Alternatively, in analogy to \vela, which
also shows a strong energy dependence of the pulse profile
\citep{kreykenbohm99a,kreykenbohm02b}, increased photoelectric
absorption or the accretion column passing through the line of sight
can also be invoked to explain the observed profiles
\citep{nagase83a}.  Due to the large and Compton thick absorbing
column during the flare ($N_{\text{H}} \sim 10^{23}\,\text{cm}^{-2}$,
see below) it seems more likely that this latter explanation holds for
\gx as well.  To investigate the causes of the pulse profile
variations, a detailed spectroscopic analysis of the pulse phase
resolved X-ray spectrum is required.

\section{Phase resolved spectra}
\label{spectra}
\subsection{Introduction}
As mentioned above, in the standard picture of accreting X-ray pulsars
the strong magnetic field of the neutron star couples to the infalling
material and channels it onto the two magnetic poles where two hot
spots emerge \citep{burnard91a}. If the magnetic axis is offset from
the rotational axis, the rotation of the neutron star gives rise to
pulsations \citep{davidson73a}.  The X-ray spectrum observed from the
accretion column and the hot spot has generally been modeled as an
exponentially cutoff power law, which is modified by strongly varying
photoelectric absorption from the accretion stream or the stellar wind
at low energies \citep{white80a,white83a}. Furthermore, an Fe
K$\alpha$ fluorescence line is observed at \ca6.4\,\keV.
\citet{swank76a} also observed an iron edge near 7\,\keV.

The strong magnetic field ($B\sim 10^{12}\,\text{G}$) at the neutron
stars' magnetic poles leads to the formation of cyclotron resonant
scattering features (CRSFs, ``cyclotron lines''). These lines result 
from resonant scattering of electrons in Landau levels in the
$\sim$$10^{12}\,\text{G}$ magnetic field of neutron star
\citep[see][and references therein]{coburn02a,araya99a,meszaros85a}.
The energy of the feature is nominally given by
 \begin{equation}
 E_{\text{C}}=11.6\text{\,\keV} \times \frac{1}{1+z} \times
 \frac{B}{10^{12}\,\text{G}} 
\end{equation}
where $E_{\text{C}}$ is the centroid energy of the CRSF, $z$ is the
gravitational redshift at the scattering site (if the scattering
occurs at the surface of the neutron star, $z\sim0.3$, depending on
the equation of state), and $B$ the magnetic field strength.  The
energy of the fundamental CRSF thus gives a direct measure of the
magnetic field strength in the line forming region.

\begin{figure}
\includegraphics[width=\columnwidth]{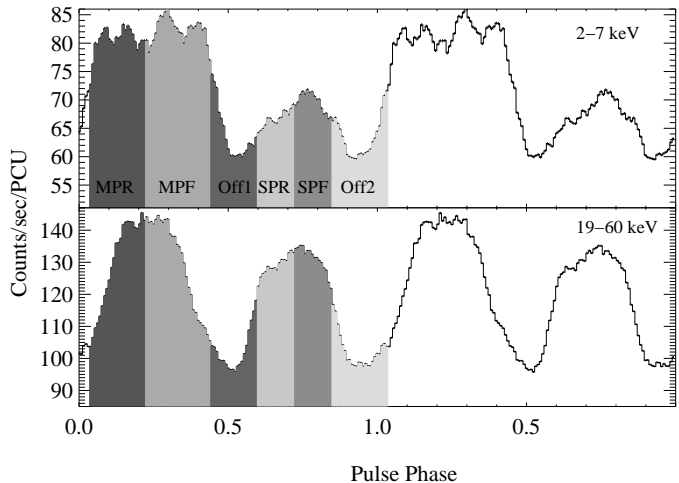}
\caption{Definition of the six phase bins used for phase resolved
  spectroscopy using pulse profiles of data taken \emph{after} the
  pre-periastron flare (see Fig.~\ref{lightcurve}). Different shadings
  show the six phase bins in use: rise and fall of the main pulse (MPR
  and MPF), rise and fall of the secondary pulse (SPR and SPF), and
  the two pulse minima (OFF1 and OFF2).  }
\label{phase_desc}
\end{figure}

Since the physical conditions are expected to vary over the emission
region, the X-ray spectrum is expected to change with the viewing
angle and therefore with pulse phase. This change is especially
important when analyzing CRSFs since their shape depends strongly on
the viewing angle \citep{araya00a,araya99a,isenberg98a}, thus making a
spectrum averaged over the whole pulse period difficult to interpret.
We therefore analyzed the data from \gx using six phase intervals as
defined in Fig.~\ref{phase_desc}: the rise and fall of the main pulse,
the rise and fall of the secondary pulse, and the two pulse minima.
These intervals were chosen to give a good coverage of the pulse with
a similar signal to noise ratio.

\subsection{The spectral model}
\label{models}

The lack of theoretical models of high enough quality for
comparison with observational data forces the use of
empirical spectral models for the characterization of
the observed continuum \citep{kreykenbohm99a}. Here, we describe the
continuum produced in the accretion column of the neutron star,
$I_{\text{NS}(E)}$, by a power law with photon index $\Gamma$ which is
cutoff by the Fermi-Dirac cutoff \citep{tanaka86a},

\begin{equation}
I_{\text{cont}}(E) = A_{\text{PL}}
\frac{E^{-\Gamma}}{\exp\left((E-\ecut)/\efold\right)+1} 
\label{fd}
\end{equation}
where $A_{\text{PL}}$ is a normalization constant, and where $\ecut$
and $\efold$ are called the cutoff and folding energy of the
Fermi-Dirac cutoff. This continuum shape is then modified by a CRSF
modeled as an absorption line with a Gaussian
optical depth profile \citep[][Eqs.~6 and 7]{coburn02a},
\begin{equation}\label{gabs}
\tau_{\text{GABS}}(E) = d_{\text{C}}
   \times  \exp\left(
 -\frac{1}{2}\left(\frac{E-E_{\text{C}}}{\sigma_{\text{C}}}\right)^2
                \right)
\end{equation}
where $E_{\text{C}}$ is the energy, $\sigma_{\text{C}}$ the width, and
$d_{\text{C}}$ the depth of the CRSF. 
The shape of this absorption line is simpler than the
pseudo-Lorentzian line shape \citep[as used by the \textsl{XSPEC}
model CYCLABS and in earlier works; see, e.g.,][]{makishima90a} and
gives equally good fits.  The overall X-ray spectrum emitted by the
neutron star is then given by
\begin{equation} 
I_{\text{NS}}(E) =  I_{\text{cont}}(E) \exp\left(-\tau_{\text{GABS}}(E)\right)
\end{equation}
where $I_\text{cont}$ is the continuum before scattering as
  given by, e.g., Eq.~\ref{fd}.  Due to the strong stellar wind of
Wray~977, however, the observed X-ray spectrum of \gx\ is strongly
modified by photoelectric absorption, especially during times when the
neutron star is close to \wray, such as in our observation. As the
material enshrouding the neutron star is most likely clumpy, we expect
(part of) the neutron star's X-rays to be strongly absorbed close to
the neutron star ($N_\text{H,2}$ in Eq.~\ref{eq:apc}). Furthermore,
all X-rays emerging from this zone will also be subject to
photoelectric absorption in the overall stellar wind of \wray
($N_\text{H,1}$ in Eq.~\ref{eq:apc}).  A model describing such a
physical situation is the \emph{absorbed partial covering model},
which has the form
\begin{equation}\label{eq:apc}
   I_{\text{obs}}(E) =  \text{e}^{-\sbf N_{\text{H},1}}\left( 1 + c
   \text{e}^{-\sbf N_{\text{H},2}} \right)
    I_{\text{NS}}(E) + I_{\text{Fe}}(E)
\end{equation}
where $\sbf$ is the bound free absorption cross section per Hydrogen
atom \citep{wilms00a}, $N_{\text{H},1}$ and $N_{\text{H},2}$ are the
column densities of the two components, and $c$ is the covering
fraction of the absorber responsible for $N_{\text{H},2}$. The
possibility of a fluorescent Fe K$\alpha$ line is accounted for by a
Gaussian emission line, $I_{\text{Fe}}$.

A second possible scenario takes into account that during periastron
passage the neutron star is extremely close to the surface of \wray. A
large fraction of the neutron star's X-rays are thus intercepted by
\wray where they are backscattered by Thomson scattering or
absorbed.  This situation is analogous to the Compton reflection hump
observed in active galactic nuclei \citep{lightman88a}. In principle,
the X-rays backscattered from within the atmosphere of \wray could be
responsible for a significant part of the total observed X-ray flux.
A model describing this physical situation has the form
\begin{equation}\label{eq:refl}
   I_{\text{obs}}(E) = \text{e}^{-\sbf N_{\text{H}}}
   \left(I_{\text{NS}}(E) + \text{R}_{I_{\text{NS}}}(E; \Omega/2\pi)
   + I_{\text{Fe}}(E) \right)
\end{equation}
where all symbols are as described above and where
$\text{R}_I(E; \Omega/2\pi)$ describes the spectrum resulting
from Compton reflecting X-rays with spectral shape $I(E)$ off 
gas with covering factor $\Omega/2\pi$
\citep{magdziarz95a}. Again, this spectrum is photoabsorbed in the
strong stellar wind of \wray.

We note that these two pictures are not exclusive: the accretion
stream might well be patchy close to the neutron star, and the X-ray
spectrum emerging from the accretion stream would then be reprocessed
in the atmosphere of \wray. Detailed spectral fitting shows, however,
that applying either of the above models already results in
$\redchi\sim 1$.  Combining the partial covering model and the
reflection model is therefore not necessary.  With the available
spectral data alone, it is not possible to test which of the two
continuum models is the more realistic one -- the absorption column of
the partial covering model is so high that it mimics the features of
the soft end of the reflection spectrum (note that the shape of the
reflection spectrum below 10\,keV is essentially identical to the
shape of the incident continuum, absorbed by a column equivalent to
one Thomson optical depth). Since the major aim of this work is the
study of the cyclotron line, which is not affected by the details of
the modeling of the continuum below 10\,keV, we will concentrate our
discussion in the remainder of this paper on the behavior of the
source as reflected in the fits of the absorbed partial covering model
(Table~\ref{fits1}). The corresponding fit parameters from the
reflection model are listed in Table~\ref{fits2}.

\begin{figure}
\includegraphics[width=\columnwidth]{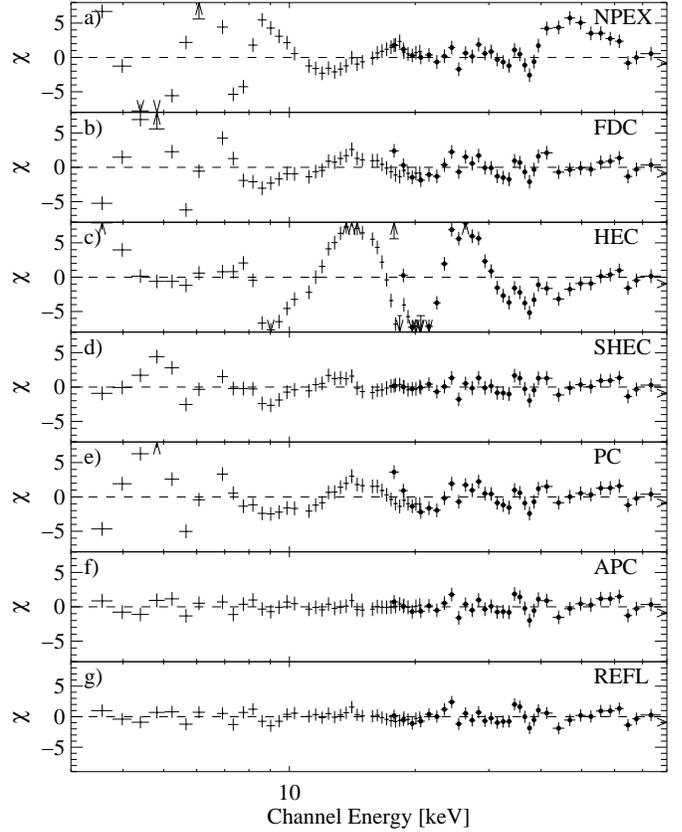}
\caption{Residuals of fits of conventional neutron star continuum
  models to phase bin SPR (see Fig.~\ref{phase_desc}). Unless
  otherwise noted, the residuals shown include photoelectric
  absorption, an additive Fe K$\alpha$ line, and the CRSF \textbf{a}:
  Negative Positive Exponential \citep[NPEX,][]{mihara95a,mihara98a},
  with strong residuals especially below 10\,\keV. Despite the
  inclusion of a CRSF at 32\,\keV, significant residuals remain above
  40\,\keV ($\redchi=17.2$ before, $\redchi=10.7$ after including the
  line).  \textbf{b}: power law with a Fermi-Dirac cutoff
  \citep[FDC,][$\redchi=9.6$ without a CRSF (not shown), 5.5 with CRSF
  at \ca35\,\keV]{tanaka86a} \textbf{c}: power law with high energy
  cutoff after \citet[][HEC; $\redchi=25.4$]{white83a}, exhibiting a
  very prominent and spurious feature. Because of this spectral
  artifact, this model should not be used for cyclotron line searches
  \citep{kretschmar96b,kreykenbohm99a}; for comparison with earlier
  observations the best fit values are shown in Table~\ref{hec}.
  \textbf{d}: the Smoothed High Energy Cutoff (SHEC), the HEC model
  smoothed by including a Gaussian absorption line (GABS) at the
  cutoff energy \citep{coburn02a}, describes the data very well after
  applying a CRSF at 35\,\keV: $\redchi=2.6$ before including a CRSF
  (not shown), and 1.7 after the inclusion; see also Table~\ref{hec},
  \textbf{e}: a standard partial covering model (PC, essentially this
  is Eq.~\ref{eq:apc}, setting $N_{\text{H},1}=0$) still shows
  significant deviations below 8\,\keV: $\redchi=11.0$ without a CRSF
  (not shown), $\redchi=5.4$ after including a 33\,\keV CRSF, which
  are remedied by using the \textbf{f}: Absorbed Partial Covering
  (APC) model of Eq.~\ref{eq:apc}: $\redchi=4.1$ without a CRSF (not
  shown), and $\redchi=0.8$ after the inclusion of a CRSF at 34\,keV.
  \textbf{g}: Reflection Model of Eq.~\ref{eq:refl}. Residuals are
  comparable to those of the APC; therefore, we use the APC model
  along with this reflection model to analyze the data. }
\label{compare_models}
\end{figure}

To allow the comparison of the spectral shape used here with other
parameterizations of the continuum, we also modeled the data with
several other standard pulsar continua (with and without a CRSF),
especially using the high energy cutoff \citep[][ see
Table~\ref{hec}]{white83a}. Residuals for these continua for the rise
of the secondary pulse are shown in Fig.~\ref{compare_models}. In this
pulse region (phase region SPR, see Fig.\ref{phase_desc}) the CRSF is
very prominent such that the data are especially suited to illustrate
the difficulties of the standard spectral models.

\begin{table}
\caption{Parameters for spectral modeling of data from phase
  bin SPR (for definition, see Fig.~\ref{phase_desc}) with a power law
  and a high energy cutoff \citep[HEC]{white83a} or a
  smoothed version of this cutoff \citep[SHEC]{coburn02a} with and
  without the inclusion of a CRSF. At lower energies all three models
  are modified by photoelectric absorption. } 
\label{hec}

\begin{tabular}{lrrr}
\hline
\hline
Parameter  & HEC & SHEC & SHEC + CRSF \\
\hline
$N_{\text{H}} [10^{22}]$  & 26.4\err{0.3}{0.2} & 17.2\err{0.8}{0.8} & 16.9\err{0.9}{0.9} \\
$\Gamma$  & 0.14\err{0.01}{0.01} & -0.32\err{0.04}{0.04} & -0.34\err{0.04}{0.05} \\
\ecut [\kevnxs]  & 19.86\err{0.02}{0.03} & 18.43\err{0.14}{0.14} & 18.22\err{0.34}{0.18} \\
\efold [\kevnxs]  & 6.08\err{0.04}{0.02} & 5.29\err{0.05}{0.05} & 5.44\err{0.09}{0.08} \\
Fe [\kevnxs]  & 6.50\err{0.03}{0.02} & 6.47\err{0.01}{0.01} & 6.46\err{0.01}{0.01} \\
Fe-$\sigma$ [\kevnxs]  & 0.25\err{0.02}{0.03} & 0.38\err{0.02}{0.02} & 0.38\err{0.02}{0.02} \\
$\tau_{\text{C}}$  & --  & --  & 0.14\err{0.05}{0.03} \\
\ecyc [\kevnxs]  & --  & --  & 34.9\err{1.7}{1.3} \\
$\sigma_{\text{C}}$  & --  & --  & 3.4\err{2.8}{1.5} \\
\hline
$\chi^2$ (dof) & 1620.4 (64) & 157.1 (61) & 99.9 (58) \\
\hline
\end{tabular}



\end{table}

\begin{figure}
\includegraphics[width=\columnwidth]{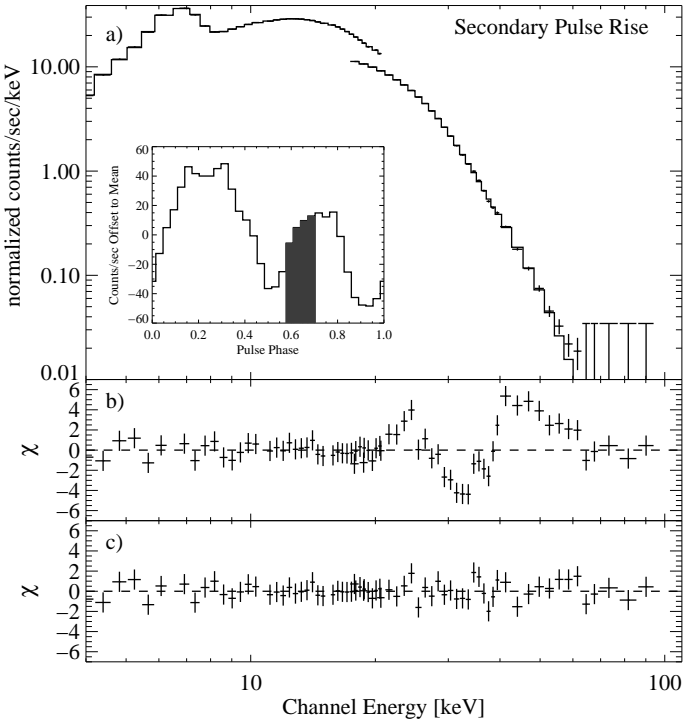}
\caption{\textbf{a}: Data and folded model of the rise of the secondary
  pulse. The model is an absorbed partial covering model (for
  discussion, see text). \textbf{b}: Residuals for the model without a
  CRSF and \textbf{c}: with a CRSF at 34.2\err{1.1}{0.9}\,\keV. The
  inset shows the pulse profile of \gx in the energy range from
  5\,\keV to 20\,\keV. The marked region is the phase bin under
  discussion -- the rise of the secondary pulse (see also
  Fig.~\ref{phase_desc}).}
\label{spr}
\end{figure}

\subsection{Spectral fits}

Since the source is extremely variable, we accumulated phase resolved
spectra not only for the whole data set, but separately also for the
pre-periastron flare and the low following the flare. With a total
usable exposure time of 176\,ks for  \pca and 89\,ks live time 
for \hexte the phase resolved spectra from the whole data set provide
excellent statistical quality (note that the exposure time in \hexte
is typically lower by \ca40\% due to the dead time of the
instrument; during our observation, the orbit of \xte was very close
to the South Atlantic Anomaly thus giving rise to an increased
particle background resulting in a deadtime of about \ca50\% in the
\hexte). In the following paragraphs, we will use this complete
dataset for our analysis. See Tables~\ref{fits1} and \ref{fits2} for a
list of the spectral parameters.

The total usable exposure time during the flare is \ca25\,ksec for
\pca and \ca12\,ksec for \hexte. Even though this is much shorter than
that of the whole observation, the flare spectra dominate over the
rest of the observation due to the much higher flux during the flare.
Since the low luminosity post periastron data have a much lower
statistical quality, the uncertainties of all parameters from this
time are fairly large and the CRSF could not be detected in any phase
region with high significance. 

Since the data from the flare are very similar to the data
  from the complete set we only discuss the spectral parameters for
  the complete data set (unless mentioned otherwise). The data from
  the extended low following the pre-periastron flare (see
  Fig.~\ref{asmlight} and Fig.~\ref{lightcurve}) have a low
  statistical quality and are therefore not discussed in detail.

\subsubsection{The X-ray spectrum below 10\,keV}
As is expected from the physical picture outlined in
Sect.~\ref{models}, the spectral parameters identified with the
stellar wind -- $N_{\text{H},1}$, $N_{\text{H},2}$, and the parameters
of the Fe K$\alpha$ line -- do not depend significantly on the pulse
phase as they originate very far away from the neutron star compared
to the accretion column where the X-rays and the pulse forms.

The high column density of the stellar wind as measured during the
whole observation, $N_{\text{H,1}}\sim 2\times
10^{23}\,\text{cm}^{-2}$ indicates how deep the neutron star is
embedded in the wind.  With a column density of as high as
$2.3\times10^{24}\,\text{cm}^{-2}$, the second component of the
partial covering model is close to being Compton thick. As explained
above, this large value is responsible for our inability to
differentiate between the partial covering model and the Compton
reflection model.

In the low following the pre-periastron flare, the absorbing columns
are measured to be even higher.  The column density for the ``weakly''
absorbed component, $N_{\text{H},1}$, varies between
$3.5\times10^{23}$ and $4.5\times10^{23}\,\text{cm}^{-2}$ and is thus
twice as large as during the flare, and also the column density
$N_{\text{H},2}$ is increased: it varies between
$4\times10^{24}\,\text{cm}^{-2}$ to over $10^{25}\,\text{cm}^{-2}$.
However, the observed count rate is not only due to the strong
photoelectric absorption but also due to an intrinsically lower source
luminosity: while the \emph{unabsorbed} flux during the flare is about
$1\times10^{-8}$\,erg\,s$^{-1}$, the \emph{absorbed} flux, however, is
$\sim4.9\times 10^{-9}$\,erg\,s$^{-1}$. During the low the unabsorbed
flux is lower by a factor of three ($\sim 3.5\times
10^{-9}$\,erg\,s$^{-1}$) and the absorbed flux by a factor of more
then six ($\sim7\times10^{-10}$\,ergs s$^{-1}$).

The high column density of the wind is also responsible for the strong
Fe K$\alpha$ fluorescence line. The line energy of
$6.48\err{0.03}{0.02}\,\keV$, corresponding to an ionization level
from \ion{Fe}{v} to \ion{Fe}{ix}, implies that the iron, and thus
probably all circumstellar material, is mildly ionized. 
The width of the line,
$\sigma_{\text{Fe}}=0.33\err{0.02}{0.02}\,\keV$, is close to the
energy resolution of the \pca, the very small uncertainties indicate
that the line width is indeed measured.

\subsubsection{The pulsar continuum and the cyclotron line}
Pulse phase spectroscopy allows us to study the variation of the
pulsar emission over the X-ray pulse. Our data analysis shows that the
photon index $\Gamma$ is almost constant over the pulse and only
varies between $-0.2\pm 0.1$ and $0.1\pm0.1$. Note that the spectrum
is slightly but systematically softer in all phase bins during the
pre-periastron flare.  The parameters of the cutoff also do not
vary over the pulse: The folding energy \efold is only variable
between 5.0\,\keV and 5.9\,\keV, while the cutoff energy \ecut is
consistent with a constant value of 15.5\,\keV except for the rise of
the main pulse where it is marginally lower
($\ecut=12.9\err{1.8}{1.6}\,\keV$).

This continuum model alone, however, does not result in an acceptable
fit in any of the phase bins (see e.g., Fig.~\ref{spr}b). In all phase
bins a large absorption line like structure is present between
30\,\keV and 40\,\keV (see Fig.~\ref{cycvar}).  After the inclusion of
a CRSF, the fits are acceptable in all phase bins with $\redchi\sim
1.0$ (see Tables~\ref{fits1} and~\ref{fits2}). The improvement is
extremely significant: according to the $F$-Test
\citep{bevington}\footnote{Note that in even if systematic
  uncertainties are not an issue, however, the extremely low false
  alarm probabilities make the detection of the line stable against
  even crude mistakes in the computation of the significance.} the
probability of a chance improvement is $2.5\times10^{-11}$ in the fall
of the secondary pulse (the bin with the smallest improvement).  We
remark that the CRSF is found at almost the same energy and the same
width (within uncertainties), but with a higher \redchi when using the
SHEC model applied in earlier publications (compare Tables~\ref{hec}
and~\ref{fits1}).

Contrary to the continuum, the depth and the position of the CRSF
(see Eq.~\ref{gabs} for definition of the CRSF parameters)
are strongly variable with pulse phase (see
Tables~\ref{fits1},~\ref{fits2}, and Fig~\ref{vary}).  This behavior
is fairly typical of most CRSF sources (e.g., \vela or
\object{Her~X-1}). We find that the line is deepest in the minimum
between the main pulse and the secondary pulse (phase bin OFF1) with
$d_{\text{C}}=0.32\err{0.19}{0.13}$, and it is also of comparable
depth in the adjacent phase bins, the fall of the main pulse (MPF,
$d_{\text{C}}=0.29\err{0.08}{0.06}$) and the rise of the secondary
pulse (SPR, $d_{\text{C}}=0.23\err{0.04}{0.03}$). While the depth of
the CRSF in these three phase bins has a lower limit of 0.20, the CRSF
is significantly less deep in the other three phase bins. At the same
time, the position of the CRSF changes from
$\ecyc=30.1\err{0.8}{0.7}\,\keV$ in the fall of the secondary pulse
and $\ecyc=31.0\err{0.6}{0.5}$\,\keV at the rise of the main pulse to
significantly higher energies in the fall of the main pulse and the
pulse minimum, where it becomes as high as $\sim 37.9\,\keV$
(Fig.~\ref{vary}).  Assuming a canonical mass of 1.4\,\Msun and a
radius of 10\,km for the neutron star, the typical gravitational
redshift amounts to \ca29\% at the surface of the neutron star, such
that the measured values imply a magnetic field between
$B=3.4\times10^{12}$\,G and $B=4.2\times10^{12}$\,G for \gx.

This change of the CRSF energy is correlated with its depth: the line
is deepest in the phase bins where the CRSF is at the highest energies
(correlation coefficient $>0.97$).  Note that this behavior is not due
to the intrinsically lower signal during the interpulse: during the
fall and the rise of the main pulse, where the data are of comparable
statistical quality (and the source has a similar flux), we find a
significant difference of the spectral parameters. We finally note
that the line
width of the CRSF $\sigma_{\text{C}}$ is always
in the range from 3\,\keV to 7\,\keV, typical for CRSFs at these
energies.

\begin{figure}
\includegraphics[width=\columnwidth]{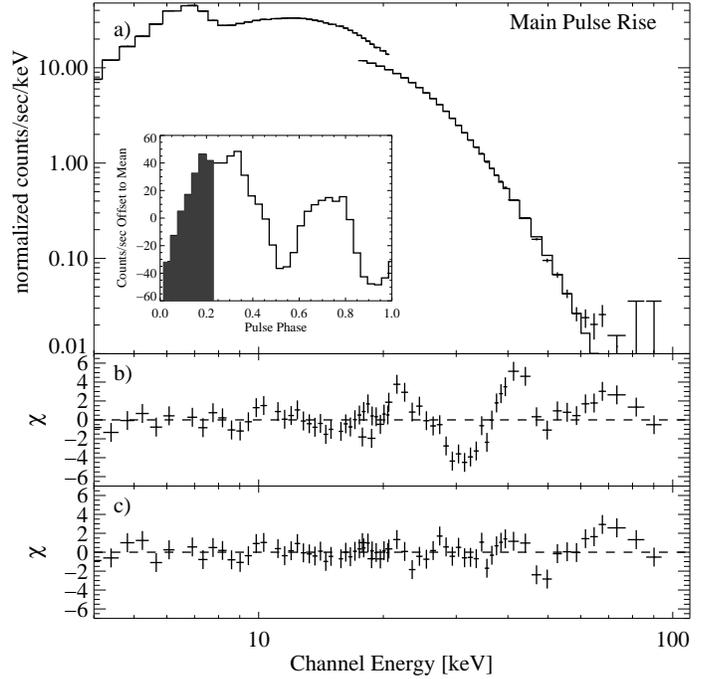}
\caption{\textbf{a} Data and folded model of the rise of the main
  pulse for the absorbed partial covering model (see text for
  discussion). \textbf{b} Residuals for a fit without a CRSF, and
  \textbf{c} with a CRSF at $31.0\err{0.6}{0.5}\,\keV$. The inset
  shows the pulse profile of \gx in the energy range from 5\,\keV to
  20\,\keV. The marked region is the rise of the main pulse. Note that
  fitting a Gaussian shaped CRSF does not completely remove the
  residuals of the CRSF; the resulting fit is acceptable, but not very
  good ($\redchi=1.2$). See text for a discussion.}
\label{mpr}
\end{figure}

\begin{figure}
\includegraphics[width=1.0\columnwidth]{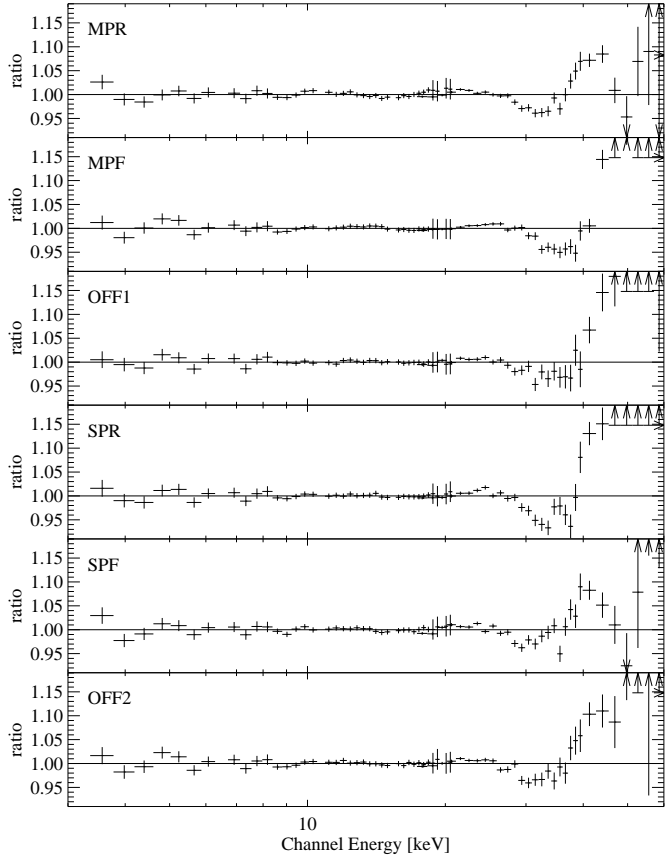}
\caption{Variation of the CRSF for the six phase bins defined in
  Fig.~\ref{phase_desc}). Note that not only the depth and the energy
  of the CRSF changes with phase but also the actual shape of the line
  itself seems also to be variable. See text for a discussion of these
  issues.
}\label{cycvar}

\end{figure}

\begin{figure}
\includegraphics[width=\columnwidth]{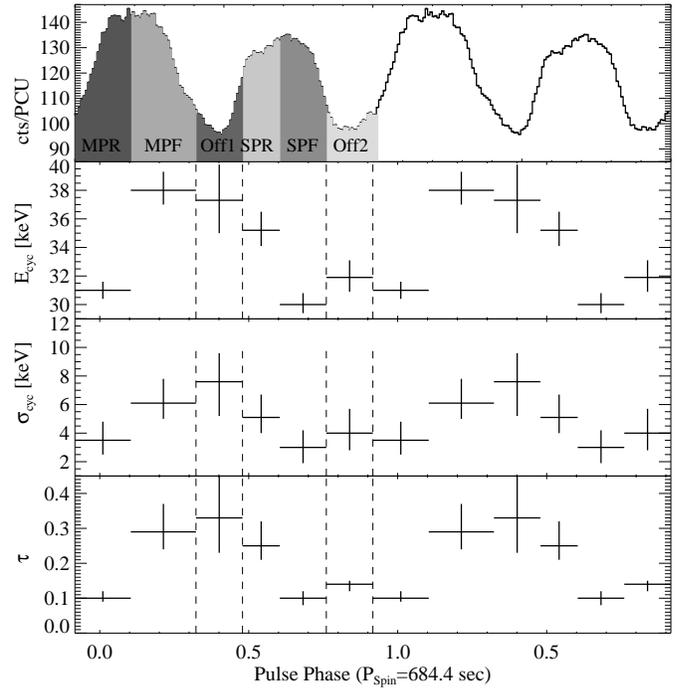}
\caption{Variation of the energy and the depth of the CRSF over the
  pulse for the APC model. Although the values are slightly different
  for the REFL model, the variation of the parameters is very
  similar. For clarity the pulse is shown twice. \textbf{a} shows the
  \pca-countrate. Note that error bars are shown, but they are too
  small to be seen in print. \textbf{b} shows the variation of the
  energy of the CRSF over the pulse. Note that the energy variation
  over the pulse is definitely non-sinusoidal and therefore not due to
  simple angle dependence. \textbf{c} shows the variation
  of the depth of the CRSF over the pulse. See text for discussion.
}
\label{vary}
\end{figure}

\begin{sidewaystable*}
\caption{Fitted parameters from pulse
  phase spectroscopy for using the whole 200\,ksec long observation of
  the periastron passage and the pre-periastron flare (see
  Fig.~\ref{lightcurve}).  The model is the absorbed partial covering
  model described in Sect.~\ref{models} by Eq.~\ref{eq:apc} with and
  without the inclusion of CRSF. Note that the inclusion of the CRSF
  between 30\,\keV and 40\,\keV in this data set improves the fits in
  all phase bins including in the pulse minima, where CRSFs are often
  either not present or insignificant \citep[see e.g.,
  ][]{kreykenbohm02b}. All uncertainties quoted in this table and
  elsewhere in this paper are 90\% confidence.  dof are the degrees
  of freedom. The flux is absorbed flux given in
  $\text{erg}\,\text{s}^{-1}$ in the 2--10\,\keV band.  }
\label{fits1}

\begin{tabular}{lrrrrrrrrrrr@{\,}rrr}
\hline
\hline
Phase & $N_{\text{H,1}} [10^{22}]$ & $N_{\text{H,2}} [10^{22}]$ & $\Gamma$ & \ecut [\kevnxs] & \efold [\kevnxs] & $E_\text{Fe}$ [\kevnxs] & $\sigma_\text{Fe}$ [\kevnxs] & $d_{\text{C}}$ & \ecyc [\kevnxs] & $\sigma_{\text{C}}$ & $\chi$ & (dof) & $\chi^2_\text{red}$& Flux \\
\hline
MPR & 22.7\err{ 2.3}{ 2.3} & 268.5\err{  8.9}{ 10.0} & -0.12\err{ 0.09}{ 0.11} &  9.9\err{ 0.7}{ 1.3} &  5.78\err{ 0.58}{ 0.58} &  6.48\err{ 0.02}{ 0.02} &  0.31\err{ 0.03}{ 0.03}& --& --& -- &    255.0 & (      62)&  4.11\\
MPR + CRSF & 22.2\err{ 1.3}{ 1.2} & 249.0\err{ 14.8}{ 20.7} & -0.05\err{ 0.13}{ 0.12} & 12.8\err{ 1.7}{ 2.0} &  5.88\err{ 0.08}{ 0.08} &  6.48\err{ 0.02}{ 0.02} &  0.33\err{ 0.03}{ 0.03} & 0.10\err{0.02}{0.01} &  31.0\err{  0.6}{  0.6} &   3.5\err{  1.3}{  1.0} &     73.9 & (      59)&  1.25 & \raisebox{1.5ex}[-1.5ex]{      1.87$\times 10^{-9}$}\\
MPF & 21.8\err{ 1.6}{ 0.5} & 265.1\err{ 16.1}{ 16.7} &  0.07\err{ 0.11}{ 0.04} & 15.8\err{ 0.7}{ 0.2} &  5.44\err{ 0.08}{ 0.04} &  6.48\err{ 0.02}{ 0.02} &  0.33\err{ 0.03}{ 0.03}& --& --& -- &    364.2 & (      62)&  5.87\\
MPF + CRSF & 22.5\err{ 1.3}{ 1.1} & 242.8\err{ 17.7}{ 20.1} &  0.06\err{ 0.10}{ 0.09} & 15.0\err{ 1.2}{ 1.1} &  5.93\err{ 0.15}{ 0.11} &  6.48\err{ 0.02}{ 0.02} &  0.34\err{ 0.03}{ 0.03} & 0.29\err{0.08}{0.05} &  38.0\err{  1.3}{  1.1} &   6.1\err{  1.7}{  1.1} &     50.4 & (      59)&  0.85 & \raisebox{1.5ex}[-1.5ex]{      1.93$\times 10^{-9}$}\\
OFF1 & 22.6\err{ 0.6}{ 0.5} & 249.6\err{ 15.5}{ 14.9} &  0.03\err{ 0.10}{ 0.06} & 16.1\err{ 0.5}{ 0.5} &  5.09\err{ 0.06}{ 0.05} &  6.49\err{ 0.02}{ 0.02} &  0.34\err{ 0.03}{ 0.03}& --& --& -- &    140.6 & (      62)&  2.27\\
OFF1 + CRSF & 22.6\err{ 1.3}{ 1.3} & 230.7\err{ 14.7}{ 13.3} & -0.18\err{ 0.26}{ 0.16} & 16.3\err{ 1.9}{ 1.6} &  5.52\err{ 0.24}{ 0.16} &  6.50\err{ 0.02}{ 0.02} &  0.34\err{ 0.03}{ 0.03} & 0.33\err{0.25}{0.10} &  37.3\err{  2.5}{  2.3} &   7.6\err{  2.0}{  2.4} &     42.4 & (      59)&  0.72 & \raisebox{1.5ex}[-1.5ex]{      1.37$\times 10^{-9}$}\\
SPR & 20.5\err{ 1.0}{ 1.0} & 218.8\err{ 14.3}{ 14.6} & -0.19\err{ 0.09}{ 0.08} & 16.2\err{ 0.4}{ 0.4} &  4.97\err{ 0.06}{ 0.05} &  6.49\err{ 0.02}{ 0.02} &  0.33\err{ 0.03}{ 0.03}& --& --& -- &    251.7 & (      62)&  4.06\\
SPR + CRSF & 20.8\err{ 1.2}{ 1.4} & 211.0\err{ 17.5}{ 18.9} & -0.21\err{ 0.10}{ 0.11} & 15.2\err{ 1.1}{ 1.1} &  5.35\err{ 0.11}{ 0.10} &  6.49\err{ 0.02}{ 0.02} &  0.33\err{ 0.03}{ 0.03} & 0.25\err{0.07}{0.04} &  35.2\err{  1.3}{  1.1} &   5.1\err{  1.6}{  1.1} &     56.0 & (      59)&  0.95 & \raisebox{1.5ex}[-1.5ex]{      1.47$\times 10^{-9}$}\\
SPF & 20.0\err{ 0.8}{ 0.6} & 234.0\err{ 12.7}{ 13.3} & -0.34\err{ 0.02}{ 0.00} & 12.8\err{ 0.8}{ 0.7} &  5.29\err{ 0.08}{ 0.06} &  6.48\err{ 0.02}{ 0.02} &  0.32\err{ 0.03}{ 0.03}& --& --& -- &    176.8 & (      62)&  2.85\\
SPF + CRSF & 20.2\err{ 1.4}{ 1.2} & 204.0\err{ 18.7}{ 21.6} & -0.25\err{ 0.11}{ 0.11} & 15.0\err{ 1.2}{ 1.1} &  5.39\err{ 0.07}{ 0.07} &  6.49\err{ 0.01}{ 0.01} &  0.33\err{ 0.03}{ 0.03} & 0.10\err{0.02}{0.02} &  30.0\err{  0.8}{  0.6} &   3.0\err{  1.2}{  1.1} &     77.8 & (      59)&  1.32 & \raisebox{1.5ex}[-1.5ex]{      1.52$\times 10^{-9}$}\\
OFF2 & 21.7\err{ 0.8}{ 0.3} & 267.0\err{ 11.8}{ 12.3} & -0.11\err{ 0.13}{ 0.06} & 13.3\err{ 1.1}{ 0.4} &  5.12\err{ 0.09}{ 0.05} &  6.50\err{ 0.02}{ 0.01} &  0.33\err{ 0.03}{ 0.02}& --& --& -- &    170.2 & (      62)&  2.74\\
OFF2 + CRSF & 22.7\err{ 1.4}{ 1.3} & 244.6\err{ 16.5}{ 19.8} & -0.03\err{ 0.11}{ 0.10} & 14.1\err{ 1.9}{ 0.8} &  5.30\err{ 0.08}{ 0.07} &  6.50\err{ 0.02}{ 0.01} &  0.34\err{ 0.03}{ 0.03} & 0.14\err{0.01}{0.02} &  31.9\err{  1.2}{  1.0} &   4.0\err{  1.7}{  1.2} &     51.0 & (      59)&  0.86 & \raisebox{1.5ex}[-1.5ex]{      1.31$\times 10^{-9}$}\\
\hline
\end{tabular}
\vspace*{2\baselineskip}

\caption{As Table~\ref{fits1}, but for the reflection model.}
\label{fits2}
\begin{tabular}{lrrrrrrrrrrr@{\,}rrr}
\hline
\hline
Phase & \nh [$10^{22}$] & $\Gamma$ & \ecut [\kevnxs] & \efold [\kevnxs] & Refl & $E_\text{Fe}$ [\kevnxs] & $\sigma_\text{Fe}$ [\kevnxs] & $d_{\text{C}}$  & \ecyc [\kevnxs] & $\sigma_{\text{C}}$ & $\chi$ & (dof) & $\chi^2_\text{red}$ & Flux\\
\hline
MPR & 21.0\err{ 0.4}{ 1.0} &  0.07\err{ 0.09}{ 0.09} & 10.4\err{ 0.7}{ 1.0} &  6.74\err{ 0.13}{ 0.08} & -3.5\err{ 0.3}{ 0.4} &  6.48\err{ 0.02}{ 0.02} &  0.35\err{ 0.04}{ 0.03}& --& --& --&  234.8 & ( 60)&  3.91\\
MPR+CRSF & 21.2\err{ 0.3}{ 1.2} &  0.09\err{ 0.09}{ 0.17} & 12.8\err{ 1.5}{ 1.6} &  6.67\err{ 0.04}{ 0.10} & -2.6\err{ 0.4}{ 0.4} &  6.47\err{ 0.02}{ 0.02} &  0.35\err{ 0.04}{ 0.03} &  0.14\err{ 1.01}{ 0.03} & 31.8\err{ 0.4}{ 0.6} &  2.2\err{ 1.2}{ 2.0}&   79.1 & ( 63)&  1.26 & \raisebox{1.5ex}[-1.5ex]{      1.87$\times10^{-9}$}\\
MPF & 22.0\err{ 1.1}{ 0.5} &  0.20\err{ 0.11}{ 0.06} & 17.2\err{ 0.3}{ 0.3} &  5.98\err{ 0.17}{ 0.07} & -1.7\err{ 0.1}{ 0.1} &  6.47\err{ 0.02}{ 0.02} &  0.36\err{ 0.03}{ 0.04}& --& --& --&  356.9 & ( 60)&  5.95\\
MPF+CRSF & 21.6\err{ 0.8}{ 0.3} &  0.13\err{ 0.12}{ 0.06} & 15.1\err{ 0.7}{ 0.4} &  6.49\err{ 0.16}{ 0.12} & -1.9\err{ 0.2}{ 0.2} &  6.47\err{ 0.02}{ 0.02} &  0.36\err{ 0.04}{ 0.03} &  0.28\err{ 0.06}{ 0.03} & 38.7\err{ 1.5}{ 0.8} &  4.7\err{ 1.5}{ 1.0}&   51.0 & ( 63)&  0.81 & \raisebox{1.5ex}[-1.5ex]{      1.93$\times10^{-9}$}\\
OFF1 & 21.7\err{ 0.9}{ 0.8} &  0.14\err{ 0.11}{ 0.07} & 16.9\err{ 0.4}{ 0.4} &  5.65\err{ 0.13}{ 0.08} & -2.2\err{ 0.2}{ 0.2} &  6.49\err{ 0.01}{ 0.02} &  0.36\err{ 0.02}{ 0.03}& --& --& --&  125.6 & ( 60)&  2.09\\
OFF1+CRSF & 21.8\err{ 0.8}{ 1.4} &  0.12\err{ 0.10}{ 0.11} & 15.6\err{ 1.0}{ 1.4} &  6.16\err{ 0.25}{ 0.25} & -2.5\err{ 0.4}{ 0.4} &  6.49\err{ 0.01}{ 0.02} &  0.35\err{ 0.03}{ 0.03} &  0.28\err{ 0.20}{ 0.08} & 38.8\err{ 3.4}{ 2.8} &  6.3\err{ 2.9}{ 2.3}&   49.7 & ( 63)&  0.79 & \raisebox{1.5ex}[-1.5ex]{      1.37$\times10^{-9}$}\\
SPR & 19.9\err{ 0.9}{ 1.1} & -0.11\err{ 0.11}{ 0.08} & 16.1\err{ 0.4}{ 0.5} &  5.57\err{ 0.12}{ 0.08} & -2.5\err{ 0.3}{ 0.2} &  6.48\err{ 0.02}{ 0.02} &  0.33\err{ 0.03}{ 0.03}& --& --& --&  222.5 & ( 60)&  3.71\\
SPR+CRSF & 19.8\err{ 0.5}{ 1.4} & -0.13\err{ 0.09}{ 0.10} & 14.8\err{ 0.8}{ 0.9} &  5.91\err{ 0.03}{ 0.14} & -2.8\err{ 0.3}{ 0.3} &  6.48\err{ 0.02}{ 0.02} &  0.33\err{ 0.02}{ 0.03} &  0.29\err{ 0.11}{ 0.05} & 35.6\err{ 1.7}{ 1.0} &  2.9\err{ 1.1}{ 1.1}&   55.5 & ( 63)&  0.88 & \raisebox{1.5ex}[-1.5ex]{      1.47$\times10^{-9}$}\\
SPF & 18.9\err{ 0.4}{ 1.3} & -0.24\err{ 0.10}{ 0.11} & 11.8\err{ 0.9}{ 0.7} &  6.08\err{ 0.08}{ 0.15} & -3.3\err{ 0.5}{ 0.3} &  6.48\err{ 0.02}{ 0.02} &  0.33\err{ 0.03}{ 0.03}& --& --& --&  144.6 & ( 60)&  2.41\\
SPF+CRSF & 19.1\err{ 0.7}{ 1.3} & -0.23\err{ 0.09}{ 0.11} & 13.3\err{ 0.8}{ 1.0} &  6.03\err{ 0.13}{ 0.14} & -2.6\err{ 0.4}{ 0.2} &  6.47\err{ 0.02}{ 0.01} &  0.33\err{ 0.02}{ 0.03} &  0.09\err{ 0.02}{ 0.02} & 32.0\err{ 0.8}{ 0.8} &  0.9\err{ 1.6}{ 0.8}&   82.9 & ( 63)&  1.32 & \raisebox{1.5ex}[-1.5ex]{      1.52$\times10^{-9}$}\\
OFF2 & 21.1\err{ 0.6}{ 1.2} &  0.08\err{ 0.09}{ 0.09} & 15.0\err{ 0.5}{ 0.5} &  5.86\err{ 0.05}{ 0.11} & -3.2\err{ 0.3}{ 0.2} &  6.49\err{ 0.02}{ 0.02} &  0.36\err{ 0.03}{ 0.03}& --& --& --&  158.5 & ( 60)&  2.64\\
OFF2+CRSF & 21.5\err{ 0.2}{ 0.2} &  0.10\err{ 0.01}{ 0.01} & 15.4\err{ 0.1}{ 0.4} &  5.96\err{ 0.02}{ 0.01} & -2.9\err{ 0.1}{ 0.0} &  6.49\err{ 0.01}{ 0.02} &  0.35\err{ 0.02}{ 0.02} &  0.14\err{ 0.04}{ 0.02} & 32.5\err{ 0.5}{ 0.4} &  3.1\err{ 0.9}{ 0.6}&   55.7 & ( 63)&  0.88 & \raisebox{1.5ex}[-1.5ex]{      1.31$\times10^{-9}$}\\
\hline
\end{tabular}

\end{sidewaystable*}

As we illustrate in Fig.~\ref{cycvar}, in addition to the changes of
the depth and position of the CRSF, there are indications for a change
in the shape of the feature itself. After fitting the CRSF in the rise
of the main pulse some residuals remain (see Fig.~\ref{mpr}),
indicating that the CRSF has a non-Gaussian line shape.  Furthermore,
the remaining residuals at \ca50\,\keV could indicate the presence of
an emission wing.  The existence of such emission wings and overall
variations of the line profile over the pulse have been predicted by
the numerical simulations of \citet{araya99a}.

We also searched for any yet undetected second CRSF. As the addition
of a second CRSF at any energy does not improve the fit, we tried to
force an additional CRSF on the best fit at twice or half the energy
of the \ca35\,\keV CRSF, especially in the phase bins where the
\redchi\ is $>1.0$ like the rise of the main pulse. A line between
15\,\keV and 20\,\keV can be excluded with very high confidence: the
upper limit for the depth of a line in this energy range is less than
$10^{-10}$. A CRSF at twice the energy (between 60\,\keV and 80\,\keV)
can not be excluded with such a high confidence, but it is still very
unlikely that there is a secondary line present; the upper limit for
the depth of such a line is 0.3, which given the low
  statistical quality of the data at 70\,\keV is very low. Typical
  line depths in this regime are expected to be \ca1.0: a CRSF with a
  depth of 0.3 and a width of \ca4\,\keV can be fitted almost
  everywhere above 70\,\keV without affecting the resulting \redchi
  significantly.  Therefore, we conclude that no other CRSF is
present in the data; we caution, however, that the source is only
weakly detected above $\sim$60\,\keV and therefore we cannot exclude
the presence of a feature in this energy range.

\section{Summary and discussion}
\label{summary}

\subsection{CRSF Variability}
The major result of this paper is the observation of a variation of
the energy of the cyclotron line by almost 8\,\keV (corresponding to
about 25\%) from $30.1\err{0.8}{0.7}\,\keV$ in the fall of the
secondary pulse to $37.9\err{1.3}{1.0}\,\keV$ in fall of the main
pulse. This large variation is similar to that found in several other
sources, such as \object{4U~1626$-$67} \citep{heindl99a} with a
variation of 25\%, \object{4U~0115$+$63} with a variation of 20\% in
the $1^{\text{st}}$ harmonic \citep{heindl99b,santangelo99b},
\object{Cen~X-3} with a variation of 30\% \citep{burderi00a}, or
\object{Her~X-1} with \ca25\% \citep{gruber01a,gruber80a}. On the
other hand, the energies of the CRSFs in \vela varies only by \ca10\%
with pulse phase \citep{kreykenbohm02b}.  Consistent with earlier
interpretations, the strong energy variation of the cyclotron line in
\gx argues that during different phases of the X-ray pulse, regions
with different magnetic fields are observed. There are several
plausible scenarios for such a suggestion.

First consider a simple scenario that we observe an accretion column
in a pure dipole field. In such a case the observed 25\% change of the
cyclotron line energy would translate into a change in the height of
the scattering region of \ca5\,km (or \ca50\% of the neutron star
radius). Such a change is rather unlikely, given that accretion
columns are generally predicted to have heights of less than 1\,km
\citep{burnard91a,becker98a,brown98a}. Apart from these theoretical
considerations, a very tall accretion column would also imply a rather
uncommon magnetic field configuration for \gx: For a $\sim$5\,km high
accretion column, the emission pattern is very likely a pure fan beam.
To entirely block the lower parts of the accretion column such that
the X-ray spectrum is dominated by the ``top'' of the column requires
that the angle between the magnetic field axis and the spin axis,
$\beta$, is at least $40^\circ$, and larger if relativistic light
bending is included.  Detailed models for the emission pattern of
XRBs, which include relativistic light propagation
\citep{kraus96a,blum00a}, as well as the statistical analysis of
accreting neutron star pulse profiles \citep{bulik03a}, however, imply
$\beta\lesssim 20^\circ$. We conclude, therefore, that the accretion
column of \gx is much lower than what would be expected if the
variation of \ecyc were due to a pure dipole field geometry.

A second possibility for the cyclotron line variability, which has
been discussed, e.g., for Cen~X-3, would be the presence of a very
wide polar cap. For a dipole configuration, \citet{burderi00a} show
that an opening angle of the order of $55^\circ$ is required to
produce a 30\% variation of the line energy. Such a large polar cap,
however, is unlikely to produce the observed X-ray pulse shape -- the
pulse profile is expected to be much broader in this case.
Furthermore, as already noted by \citet{burderi00a}, such large polar
caps would imply that the line energy varies symmetrically about the
pulse maximum, which is not the case for any of the pulsars with a
strong variation of the line energy. Finally, we also note that the
strong magnetic field confines the accreting plasma, resulting in a
much smaller opening angle of the neutron star's hot spot
\citep{litwin01a,becker98a,basko76a}. As a result we conclude that the
extended polar cap scenario is an unlikely explanation for the large
variation of the energy of the CRSF.

A third and final scenario for the pulse variation is that the
observed X-rays come from an accretion mound, as suggested, e.g., by
\citet{burnard91a}. In this case the observed variation of the line
energy is mainly due to the higher multipole components of the
magnetic field in the mound and the X-ray emissivity profile is a
combination of a fan and a pencil beam. 

Parenthetically, we note that slight variations of the line energy are
also predicted for the case of homogeneous magnetic fields, as the
location, depth, and shape of the CRSF are predicted to change with
viewing angle, i.e., pulse phase
\citep{meszaros85a,isenberg97a,araya99a,araya00a}.  The strongest of
these effects are the emission wings of the fundamental CRSF caused by
``photon spawning'' \citep{araya99a,araya00a}, where electrons excited
into a higher Landau level by resonant scattering of a photon, decay
into the ground state by emitting photons at an energy which is
roughly that of the fundamental CRSF. As a result the fundamental line
is predicted to be shallower and have a more complex shape than the
higher harmonics.  When folded through the \hexte response matrix,
these variable complex line profiles can result in slight changes of
the measured line energy.  These changes, however, are much smaller
than the 25\% energy variation seen here, and can thus be excluded as
the cause for the observed variation.  Nevertheless, the inspection of
the residuals of the pulse phase resolved fits indicates a possible
change of the shape of the CRSF over the pulse (Fig.~\ref{cycvar}).
Similar to, e.g., 4U~0115$+$63 \citep{heindl99b,santangelo99b}, the
presence of an unresolved complex structure of the fundamental could
be postulated to explain the not fully satisfactory $\chi^2$ values in
some phase bins (e.g., $\redchi\ca 1.2$ for bin MPR (see
Fig.~\ref{mpr}), as compared to $\redchi<1.0$ in the other phase
bins).  With the increasing energy resolution of modern gamma-ray
detectors, it is foreseeable that such profiles will be resolvable
with newer instruments such as, e.g., the SPI instrument on INTEGRAL.

\subsection{Implications of the Line variability}
In recent years, several groups have searched for correlations between
the parameters of the CRSF and the X-ray continuum in the hope of
deducing information about the line and continuum formation process
\citep[][and references therein]{makishima99a,coburn01b}.  A major
disadvantage of these searches, however, has been that they typically
used pulse \emph{averaged} data instead of pulse resolved
spectroscopy, complicating the interpretation of the correlations
found.

\begin{figure}
\includegraphics[width=\columnwidth]{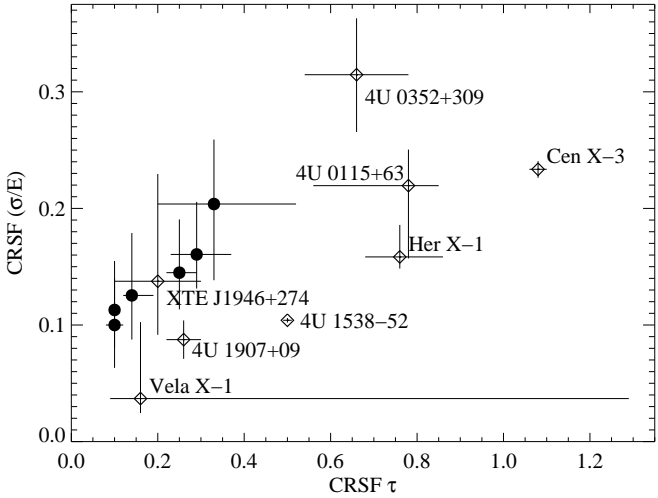}
\caption{Fractional CRSF width $\sigma_\text{C}/\ecyc$ versus the
  depth of the CRSF for several accreting neutron stars from \xte
  data. Diamonds: values derived by \citet{coburn02a} from phase
  averaged spectra.  Filled circles: values derived from phase
  resolved spectra for \gx (this work).}
\label{depth_ratio}
\end{figure}

One of the most interesting new correlations found in the large sample
of X-ray pulsars observed with \xte and analyzed in a uniform manner
by \citet{coburn01b} and \citet{coburn02a} has been between the
relative width of the CRSF, $\sigma_\text{C}/E$, and its depth,
$d_\text{C}$. As noted by \citet{coburn02a}, part of the
$\sigma_\text{C}/E_\text{cyc}$--$d_\text{C}$-plane is inaccessible to
\xte due to observational constraints; however, our data all fall into
the range that is well observable. As shown in Fig.~\ref{depth_ratio},
our phase resolved results for \gx confirm the overall correlation:
deeper CRSFs are generally broader. We note that this agreement
between the phase averaged and the phase resolved data suggests that
the correlations found by \citet{coburn02a} from the phase averaged
data are indeed real and not due to effects of the averaging.

Having many data points for one object helps in understanding the
$\sigma_\text{C}/E_\text{cyc}$-$d_\text{C}$ correlation in terms of a
physical model.  As we have outlined in the previous section, it is
likely that the X-rays observed from \gx are produced in an accretion
mound of moderate height at the magnetic poles of the neutron star.
Consider first the case that only one pole is visible. Over the X-ray
pulse we view this pole under different viewing angles, $\theta$,
where $\theta$ is the angle between the line of sight and the magnetic
field in the accretion column.  For the high temperatures of the
accretion column, the line width is expected to be dominated by
thermal broadening.  As shown, e.g., by \citet{meszaros85a}, the
anisotropic velocity field of the electrons in the accretion column
leads to fractional line widths of
\begin{equation}\label{eq:broad}
\frac{\sigma_\text{C}}{\ecyc} \propto \sqrt{\frac{kT_\text{e}}{m_\text{e} c^2}
  \cos^2\theta}
\end{equation}
where $kT_\text{e}$ is the temperature of the electrons along the
magnetic field lines. Basic Comptonization theory suggests that
$kT_\text{e}$ can be estimated from the folding energy of the pulsar
continuum, $\efold$ \citep[e.g.,][and references therein]{burderi00a}.
If we furthermore assume that the seed photons for the Compton
upscattering in the accretion column are created throughout the volume
of the accretion column, then detailed Monte Carlo simulations, e.g.,
of \citet{isenberg97a} show that the depth $d_\text{C}$ of the CRSF is
expected to be largest for a slab geometry when the line of sight is
almost perpendicular to the direction of the magnetic field.  For a
uniform temperature accretion column, where $kT_\text{e}$ is constant,
these slab models thus predict an anti-correlation between
$d_\text{C}$ and $\sigma_\text{C}/\ecyc$, in contradiction to the
observation.  On the other hand, assuming a cylindrical geometry, the
depth $d_\text{C}$ of the CRSF is expected to be largest when the line
of sight is parallel to the magnetic field
\citep[][Fig.~19]{isenberg97a}. This geometry therefore predicts a
correlation between $d_\text{C}$ and $\sigma_\text{C}/\ecyc$, in good
agreement with the observation.

We note, however, that even phase resolved data are difficult to
interpret. In the above discussion, we assumed that the observed data
are dominated by emission from one homogeneous emission region. In
reality, this is not the case and the observed data are generally a
mixture of contributions from both magnetic poles, which could
influence the observed correlation. While it is difficult to
disentangle the contributions from the two poles
\citep{kraus96a,blum00a}, it seems likely that the physical conditions
at both poles are slightly different. We would then expect the
parameters of the X-ray continuum emitted by each pole to be
different, which would be reflected by changes in the observed
continuum parameters. In \gx, however, the continuum parameters as
exemplified by $\efold$ are remarkably stable, suggesting either that
only one magnetic pole contributes to the observed data, or that the
poles have similar temperature. For this reason, it seems unlikely
that a mixture of flux from the two poles causes the observed
$\sigma_\text{C}/d_\text{C}$ correlation.  Given the
  contradiction of predicted spectra of a slab geometry with the
  observations, the slab geometry can either be ruled out or
  additional physical processes are operating.

\begin{figure}
\includegraphics[width=\columnwidth]{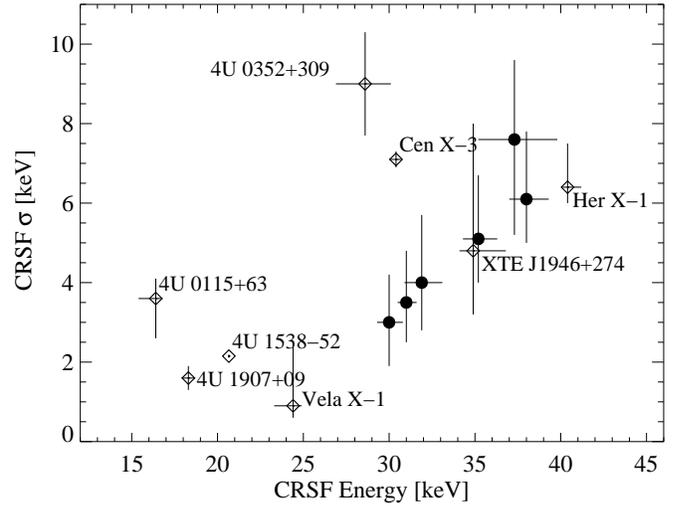}
\caption{CRSF width $\sigma_\text{C}$ versus CRSF Energy $E_\text{C}$
  for the same sample of accreting neutron stars as in
  Fig.~\ref{depth_ratio}. Note the very strong correlation between
  \ecyc and $\sigma_\text{C}$ for \gx (correlation coefficient 0.93;
  filled circles).}
\label{energy_sigma}
\end{figure}

Finally, we note a second correlation found in our phase resolved data
that is also present in the phase average data \citep{coburn02a}: a
correlation between the line width $\sigma_\text{C}$ and the energy of
the line \ecyc\ shown in Fig.~\ref{energy_sigma}. In terms of the
simple cyclotron line broadening theory of Eq.~\ref{eq:broad} this
correlation is rather unexpected since such a strong correlation (0.93
for \gx) is only possible if $\cos\theta$ does not vary appreciably or
$\theta$ is close to $90^\circ$, where a relativistic treatment of
Eqn.~\ref{eq:broad} is required and $\sigma_\text{C}$ becomes
independent of $\cos \theta$. A variation of $\theta$ is required,
however, since strong X-ray pulses are observed.  For theta close to
$90^\circ$, however, we would not expect to see strong pulsations,
contrary to what is observed. We therefore think that smaller values
of theta are more likely.  As an example, for $\theta=60^\circ$, the
correlation constrains the viewing angle to vary by $\pm 6^\circ$
around its mean value, tight even in the case of a pure fan beam.

\subsection{Summary}
To summarize, our analysis of the \xte data of \gx shows
\begin{enumerate}
\item The continuum of \gx is well described by an absorbed and
  partially covered pulsar continuum model. An alternative explanation
  for the X-ray spectrum is a reflected and absorbed pulsar continuum.
  Current data do not allow us to distinguish between these
  alternatives.
\item \gx shows a strong cyclotron line at an energy of \ca35\,keV.
  The line energy, depth, and width are variable with the pulse phase
  and do not depend on the chosen continuum model.
\item The variability of the CRSF energy as well as the pulse profile
  suggest that the X-rays from \gx originate in an accretion mound,
  and that the change in energy is not due to a change in height in a
  dipole field.
\item The correlation between the relative width of the line and its
  depth is suggestive of changes in the angle between the line of
  sight and the magnetic field at the neutron star pole, in agreement
  with the standard paradigm for pulsating X-ray sources, but
  contradiction with the numerical simulations for the depth of the
  line for certain geometries.
\end{enumerate}

\begin{acknowledgements}
  This work has been financed by DLR grant 50OR0002, DAAD grants for
  the collaboration with UCSD and the Universidad di Alicante, NASA
  grants BAS5-10729 and NAS5-30720, and NSF travel grant
  NSF-INT-0003773.
\end{acknowledgements}

\bibliographystyle{aa}
\bibliography{mnemonic,aa_abbrv,velax1,div_xpuls,xpuls,cyclotron,books,roentgen,satelliten,foreign,jwadd}

\end{document}